\documentclass[superscriptaddress,
amsmath, ,nofootinbib,amssymb,12pt]{revtex4-2}
\usepackage[utf8]{inputenc}
\usepackage[T2A]{fontenc}
\usepackage[english]{babel}
\usepackage{amsmath, amssymb, amsfonts}
\usepackage{xcolor}
\usepackage{geometry}
\usepackage{graphicx}
\usepackage{comment}
\usepackage{caption}
\usepackage{lipsum}
\usepackage[normalem]{ulem}
\usepackage{overpic}
\usepackage{slashed}
\usepackage{epsfig}

\numberwithin{equation}{section}

\begin{document}
\renewcommand{\refname}{References}
\renewcommand{\figurename}{Fig.}

\title{The Non-Relativistic Limit of Five-Dimensional Boson Stars and Scale Invariance Violation}

\author{Eduard Kim}
\affiliation{Institute for Nuclear Research of RAS, prospekt 60-letiya Oktyabrya 7a, Moscow, 117312}
\affiliation{Moscow Institute of Physics and Technology, Institutsky lane 9, Dolgoprudny, Moscow region, 141700}

\author{Oleg Kondrashov}
\affiliation{Lomonosov Moscow State University Branch in the City of Sarov, Sarov, 607328}

\author{Emin Nugaev}
\affiliation{Institute for Nuclear Research of RAS, prospekt 60-letiya Oktyabrya 7a, Moscow, 117312}
\affiliation{Lomonosov Moscow State University Branch in the City of Sarov, Sarov, 607328}

\author{Yakov Shnir}
\affiliation{BLTP JINR, Joliot-Curie St 6, Dubna, Moscow region, 141980}

\begin{abstract}
We investigate the non-relativistic limit of spherically symmetric boson stars in $(4+1)$-dimensional general relativity (GR). In this limit, the resulting Newtonian theory exhibits an exact scale invariance, leading to a critical boson star configuration with a fixed global $U(1)$ charge and vanishing binding energy. We identify the origin of these properties in the restored scale symmetry and analyze how relativistic and gravitational corrections break this symmetry. By comparing the full (4+1)-dimensional Einstein–Klein–Gordon system with the  Wick–Cutkosky model, we quantify the effects of these corrections on the energy–charge relation and the kinematic stability of the configurations. Our results demonstrate how weak violation of scale invariance lifts the degeneracy of the non-relativistic solutions and determines the behavior of boson stars away from the critical limit.

\end{abstract}
\maketitle

\emph{Dedicated to the memory of Dmitry Levkov
}
\section*{Introduction}

Boson stars (BS) are localized, self-gravitating configurations of bosonic matter that were introduced in pioneering works \cite{Kaup:1968zz,Feinblum:1968nwc,Ruffini:1969qy}. Furthermore, these non-perturbative objects have been studied in many physical contexts, they appear as an important model for investigating various interplays between gravity, quantum fields, and compact objects, see, for example, Refs. \cite{Schunck:2003kk,Liebling:2012fv,Shnir:2022lba} for reviews.

Boson stars arise in a variety of models. Important
examples are microscopic BSs in the massive Einstein-Klein-Gordon theory
\cite{Kaup:1968zz,Feinblum:1968nwc,Ruffini:1969qy,Colpi:1986ye,Deppert:1979au},
solitonic BSs  in the models with self-interacting potentials
\cite{Friedberg:1986tq,Colpi:1986ye,Kleihaus:2005me,Kleihaus:2007vk},
Proca stars \cite{Brito:2015pxa,Sanchis-Gual:2017bhw} and axion stars in the models with periodic scalar potentials
\cite{Preskill:1982cy,Kolb:1993hw,Kolb:1993zz,Levkov:2016rkk,Guerra:2019srj}. Boson stars
have been considered as possible candidates for
dark matter \cite{Kusenko:2005du,Mielke:2002bp,Chen:2020cef} and
black hole mimickers \cite{Guzman:2009zz,Olivares:2018abq}.
These compact objects tend to delocalize in the weak non-linearity limit. 
For dense BS effects of self-interaction could be more relevant compared to gravity. In this case, compact objects could be considered as non-topological solitons, the Q-balls \cite{Coleman:1985ki}.

An interesting line of research concerns the study of boson stars in models
with extra spatial dimensions, which appear naturally in Kaluza--Klein theories,
supergravity, string theory, the braneworld scenarios or modified gravity (such as Gauss-Bonnet gravity \cite{Brihaye:2013zha}).
Boson stars in $(4+1)$ dimensions have attracted considerable attention
as simple models for understanding how higher dimensions influence gravitational equilibrium,
stability, and the structure of compact objects \cite{Prikas:2004fx,Hartmann:2010pm,Brihaye:2013zha,Brihaye:2015jja}.

Many properties of self-gravitating systems can be understood without solving the full set of
Einstein equations coupled to matter fields, but rather by using the Newtonian and first post-Newtonian (1PN)
approximations \cite{Weinberg:1972,Pissson:2014}. In particular, Newtonian approximation to boson stars in $(3+1)$ dimensions
was discussed in \cite{Silveira:1995dh,Chavanis:2011zi,Levkov:2018kau}.
The first post-Newtonian approximation  in $(4+1)$ Kaluza--Klein theory with a compact extra dimension
was studied in \cite{Xu:2007dc}, the multipole moments related to the post-Newtonian approximation
were discussed recently \cite{Heynen:2023sin}. Also, the Newtonian limit of boson stars in $(d+1)$ dimensional spacetime
was investigated in \cite{Kan:2016xkn}.

It was discussed in \cite{Brihaye:2015jja,franzin2024}, that minimal boson stars in $(4+1)$ dimensions are kinematically unstable (see Fig.\ref{fig:bin_diagrams}), i.e. fission into state of free particles is more favorable in terms of energy balance. In both papers, the authors have noticed a peculiar behavior of the BSs in the Newtonian limit. As we will show, Newtonian gravity in $(4+1)$ dimensions is scale-invariant. This symmetry has been found to be crucial for non-stationary analyses in various physical models \cite{Duval:2024eod}. The presence of scale invariance and the distinct patterns of this symmetry violation allow studying properties of different stationary field lumps in $(2+1)$ dimensions \cite{Galushkina:2025yce}.

The stability of boson stars changes drastically when self-interaction terms for the bosonic matter field \cite{marks2026} are included in the model. Additional non-linearities are defined by specific details of ultraviolet completion of the model, yet the leading self-interaction order is quartic and thus preserves the non-relativistic scale invariance in Newtonian limit \cite{Galushkina:2025hur}.

In this paper, we study spherically symmetric BSs in $(4+1)$ dimensional general relativity. The Newtonian limit for BSs is considered in Sec. \ref{chapter 1}. The discussion of symmetries and various properties of compact solutions in resulting Newtonian model is provided in Sec. \ref{chapter 2}. Sec. \ref{chapter 3} is devoted to studying different patterns of scale invariance violation. In order to capture the role of relativistic and gravitational corrections, we compare the numerical solutions in Wick-Cutkosky model and original full GR theory. We conclude with a discussion of the
results and final remarks.

\section{Newtonian Boson Stars} \label{chapter 1}
We consider the action for a massive complex scalar field minimally coupled to gravity. In general form it can be written as\footnote{Hereafter, Greek indices are spacetime indices taking values from 0 to 4, whereas Latin symbols denote spatial Euclidean indices and range from 1 to 4. Throughout the paper we use Einstein's summation rule for repeated indices.}
:
\begin{equation}
    S=\int d^5x\sqrt{-g}\left(-\partial_\mu\varphi^*\partial^\mu\varphi-m^2\varphi^*\varphi+\frac{R}{16\pi G_5}\right),
    \label{eq:1}
\end{equation}
where $G_5$ is the five-dimensional gravitational constant which is proportional to $M_{5Pl}^{-3}$. $M_{5Pl}$ is the five‑dimensional Planck mass and is the only ultraviolet (UV) cutoff scale. The introduction of self-interaction terms of mass dimension greater than $[m]^5$ automatically leads to the appearance of new UV scales \cite{Weinberg:1979, burgess2021}. $R$ is the Ricci scalar, $g = \det g_{\mu\nu}$. Further in this article, the natural system of units $\hbar = c = 1$ is chosen.
We also use Minkowski spacetime metric with the signature
\[
    \eta_{\mu\nu}=\text{diag}\left(-1,+1,+1,+1,+1\right).
\]
Due to the choice of the Minkowski metric signature, generally upper spatial indices coincide with lower ones. 

\begin{figure}[htbp]
    \centering
    \includegraphics[width=0.95\textwidth]{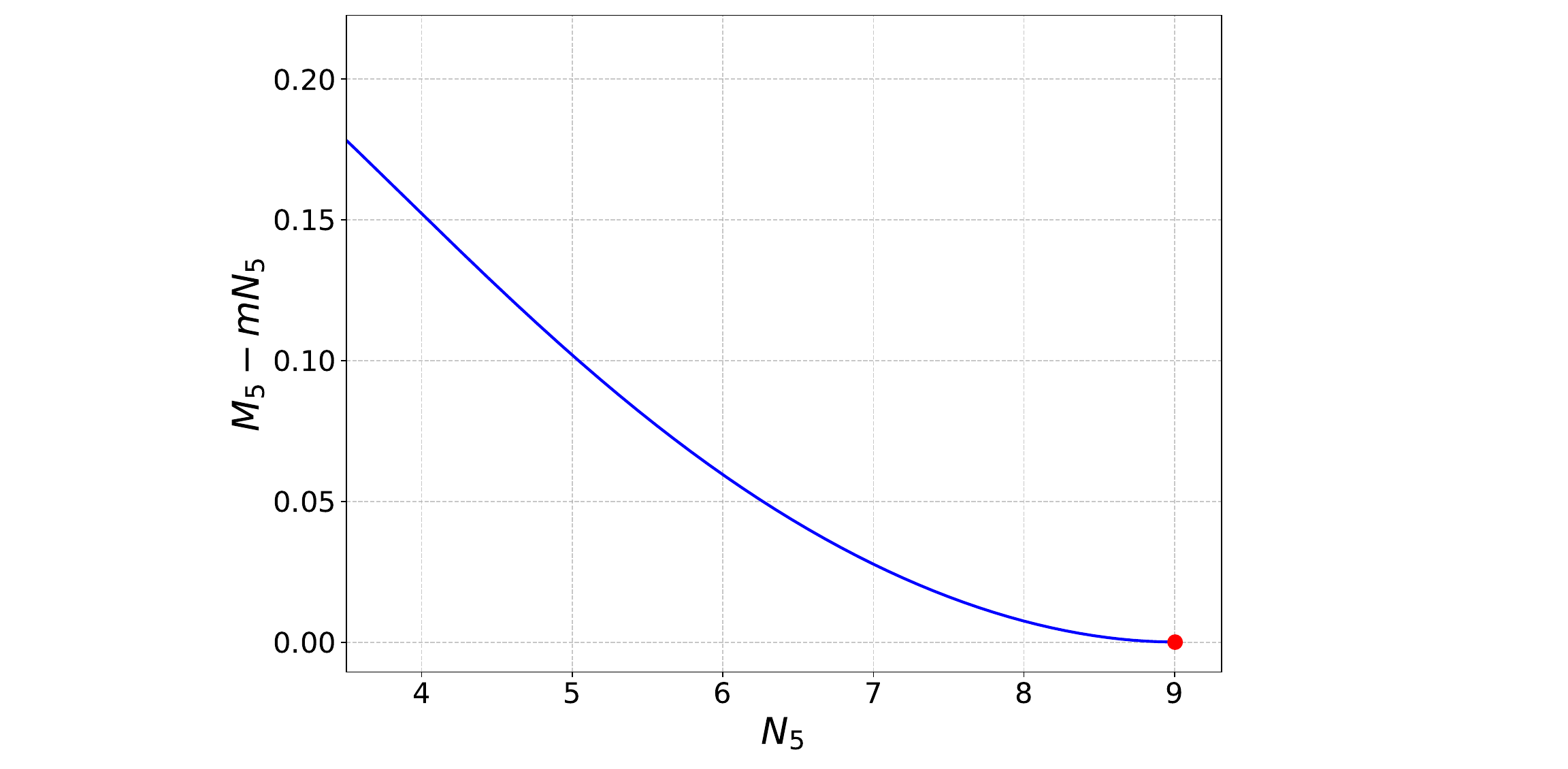}
    \caption{
    The binding energy of the configuration plotted versus the global charge of five-dimensional BS. The dot denotes the Newtonian limit of BSs with restored scale invariance. The details of the small $N_{5}$ region are omitted in this diagram because they capture strong gravity (and purely relativistic) regime.
    }
    \label{fig:bin_diagrams}
\end{figure}

The action (\ref{eq:1}) is invariant under transformation of global $U(1)$ symmetry. Thus, it possesses conserved Noether charge $N_{5}$ that corresponds to the particle number in quantized theory. For a given global $U(1)$ charge the compact object with minimal action is described by ansatz $\phi=e^{-i\omega t}f(\vec{x})$, where $\omega$ is a free frequency parameter of the solution \cite{Kaup:1968zz, Ruffini:1969qy}. This configuration represents coherent state of boson particles with mean squared velocity $\bar{v}^{2}$. 

Following recent paper \cite{franzin2024}, one can numerically study minimal boson stars in $(4+1)$ dimensions with the corresponding reparametrization choice $M_{5Pl}=m=1$. In Fig.\ref{fig:bin_diagrams}, we depict the behavior of the BS mass $M_{5}$ composed of bosons with mass $m$ and the particle number $N_{5}$ conserved due to global $U(1)$ symmetry (see App. \ref{Integral characteristics}). In order to demonstrate the instability of the minimal BSs, it is worth to consider binding energy quantity $(M_{5}-mN_{5})$ versus the particle number. A noticeable features of the diagram above: a) the binding energy has a positive sign and the BS is not a bound state; b) one can see that $M_{5}-mN_{5}=0$ in the upper limit on the global $U(1)$ charge $\text{max}\{N_{5}\}$. As we argue in this paper, both features are consequences of the scale invariance restoration in the Newtonian gravity limit. The study of Newtonian BSs requires construction of the corresponding effective theory for action \eqref{eq:1}.

\subsection{Weak gravity limit and 1PN corrections}

Let us consider the weak-field limit for gravity to investigate Newtonian BS-type solutions. 
For a systematic derivation of the weak gravity effective theory for the Einstein equations, a small expansion parameter needs to be defined (discussed in detail in \cite{Weinberg:1972,Pissson:2014}).
Indeed, for weak gravitational fields
and low energies, it is natural to choose the system's kinetic energy—or, by analogy with classical mechanics,
the characteristic squared velocity $\bar{v}^2$—as such parameter. 
This quantity should be of the same order as the characteristic Newtonian potential of the system 

\[
\frac{G_5M}{\bar{r}^2}\sim\bar{v}^2,
\]
where $M$ is the mass of the system, $\bar{r}$ is the characteristic size of the system. 

In the non-relativistic limit, the left-hand side of Einstein's equations can be derived from an expansion of the metric tensor
(for details see App. \ref{appendix a})

\begin{equation}
\left\{
\begin{aligned}
 & g_{00}=-1+\overset{(2)}{g_{00}}+\overset{(4)}{g_{00}} + \dots, \\
 & g_{ij}=\delta_{ij}+\overset{(2)}{g_{ij}}+\overset{(4)}{g_{ij}} + \dots
\end{aligned}
\right.
\label{eq:5}
\end{equation}
where the symbol $\overset{(N)}{g_{\mu\nu}}$ $\sim\bar{v}^N$.

This form of the expansion is natural for spherically-symmetric configurations. The full justification of this expansion will follow from the subsequent derivation of a self-consistent solution of the Einstein equations of motion.

The inverse metric tensor is determined by the relations:
\begin{equation}
\left\{
\begin{aligned}
 & g^{i\mu}g_{0\mu}=g^{i0}g_{00}+g^{ij}g_{0j}=0, \\
 & g^{0\mu}g_{0\mu}=g^{00}g_{00}+g^{0i}g_{0i}=1, \\
 & g^{i\mu}g_{j\mu}=g^{i0}g_{j0}+g^{ik}g_{jk}=\delta_{ij}.
\end{aligned}
\right.
\label{eq:6}
\end{equation}
We also assume:
\begin{equation}
\left\{
\begin{aligned}
  g^{00} &= -1 + \overset{(2)}{g^{00}} + \overset{(4)}{g^{00}} + \dots, \\
  g^{ij} &= \delta^{ij} + \overset{(2)}{g^{ij}} + \overset{(4)}{g^{ij}} + \dots
\end{aligned}
\right.
\label{eq:7}
\end{equation}
The relations between the components of the expansion of the metric tensor are derived from \eqref{eq:5}, \eqref{eq:6} and \eqref{eq:7}.
\begin{equation}
    \overset{(2)}{g^{00}}=-\overset{(2)}{g_{00}},\quad\overset{(2)}{g^{ij}}=-\overset{(2)}{g_{ij}},\quad\overset{(4)}{g^{00}}=-\overset{(4)}{g_{00}}-\left(\overset{(2)}{g_{00}}\right)^2.
\end{equation}
The expansion of affine connection is similar to \eqref{eq:5} and has the form:
\[
\left\{
\begin{aligned}
  \Gamma^{i}_{00}=\overset{(2)}{\Gamma^{i}_{00}}+\overset{(4)}{\Gamma^{i}_{00}}+\dots, \\
  \Gamma^{i}_{0j}=\overset{(3)}{\Gamma^{i}_{0j}}+\overset{(5)}{\Gamma^{i}_{0j}}+\dots,
\end{aligned}
\right.
\]
where the symbol $\overset{(N)}{\Gamma^{\mu}_{\nu\lambda}}\sim\bar{v}^N$.

The expansions of the components $\Gamma^i_{jk}$ and $\Gamma^0_{0i}$ are analogous to that of $\Gamma^i_{00}$. For $\Gamma^0_{00}$ and $\Gamma^0_{ij}$, the relation written for $\Gamma^i_{0j}$ holds.

Based on the expansion of the metric tensor and the affine connection in the small parameter $\bar{v}^2$, we can write the explicit form of the components $\Gamma^\lambda_{\mu\nu}$ required for further calculations.
\begin{equation}
    \left\{
\begin{aligned}
  & \overset{(2)}{\Gamma^i_{00}}=-\frac{1}{2}\partial_i\overset{(2)}{g_{00}},\\
  & \overset{(2)}{\Gamma^i_{jk}}=\frac{1}{2}\left(\partial_k\overset{(2)}{g_{ij}}+\partial_j\overset{(2)}{g_{ik}}-\partial_i\overset{(2)}{g_{jk}}\right),\\
  & \overset{(2)}{\Gamma^0_{0i}}=\overset{(2)}{\Gamma^0_{i0}}=-\frac{1}{2}\partial_i\overset{(2)}{g_{00}}, \\
  & \overset{(4)}{\Gamma^i_{00}}=\partial_0\overset{(3)}{g_{i0}}-\frac{1}{2}\partial_i\overset{(4)}{g_{00}}+\frac{1}{2}\overset{(2)}{g_{ij}}\partial_j\overset{(2)}{g_{00}}, \\
  & \overset{(3)}{\Gamma^i_{i0}}=\frac{1}{2}\partial_0\overset{(2)}{g_{ii}}.
\end{aligned}
\right.
\label{eq:9}
\end{equation}
The expressions in \eqref{eq:9} take this form because there is no distinction between upper and lower spatial indices and agree with the notation in \cite{Weinberg:1972}. 

Analogously to \eqref{eq:5}, the Ricci tensor is expanded as
\begin{equation}
        \left\{
\begin{aligned}
 & R_{00}=\overset{(2)}{R_{00}}+\overset{(4)}{R_{00}}+\dots,\\
 & R_{ij}=\overset{(2)}{R_{ij}}+\overset{(4)}{R_{ij}}+\dots,
\end{aligned}
\right.
\label{eq:10}
\end{equation}
where $\overset{(N)}{R_{\mu\nu}}\sim\bar{v}^N$.

Equations \eqref{eq:3}, \eqref{eq:9} and \eqref{eq:10} yield explicit expressions for the components of the Ricci tensor expansion:
\begin{equation}
\left\{
\begin{aligned}
  & \overset{(2)}{R_{00}}=-\frac{1}{2}\partial^2_i\overset{(2)}{g_{00}}, \\
  & \overset{(2)}{R_{ij}}=\frac{1}{2}\left(\partial_m\partial_j\overset{(2)}{g_{mi}}+\partial_m\partial_i\overset{(2)}{g_{mj}}-\partial_m^2\overset{(2)}{g_{ij}}+\partial_j\partial_i\overset{(2)}{g_{00}}-\partial_j\partial_i\overset{(2)}{g_{mm}}\right),
\end{aligned}
\right.
\label{eq:11}
\end{equation}
as well as the component $\overset{(4)}{R_{00}}$:
\[
\overset{(4)}{R_{00}}=-\frac{1}{2}\partial^2_m\overset{(4)}{g_{00}}+\partial_m\partial_0\overset{(3)}{g_{m0}}+\frac{1}{2}\partial_m\overset{(2)}{g_{mj}}\partial_j\overset{(2)}{g_{00}}+\frac{1}{2}\overset{(2)}{g_{mj}}\partial_m\partial_j\overset{(2)}{g_{00}}-
\]
\begin{equation}
  -\frac{1}{2}\partial_0^2\overset{(2)}{g_{mm}}-\frac{1}{4}\partial_k\overset{(2)}{g_{00}}\partial_k\overset{(2)}{g_{mm}}-\frac{1}{4}\Bigl(\partial_m\overset{(2)}{g_{00}}\Bigr)^2.
  \label{eq:12}
\end{equation}

Next, the right-hand side of the Einstein equations is treated \eqref{eq:2}. We expand the energy-momentum tensor in the small parameter $\bar{v}^2$:
\[
\left\{
\begin{aligned}
  T_{00} &=\overset{(0)}{T_{00}}+\overset{(2)}{T_{00}}+\dots, \\
  T_{ij} &=\overset{(2)}{T_{ij}}+\overset{(4)}{T_{ij}}+\dots,
\end{aligned}
\right.
\]
where $\overset{(N)}{T_{\mu\nu}}\sim \left(M/\bar{r}^3\right)\bar{v}^N$.
For further convenience, the notation $S_{\mu\nu}=T_{\mu\nu}-\frac{1}{3}g_{\mu\nu}T^\rho_\rho$ is introduced. The tensor $S_{\mu\nu}$ is expanded in the same way as $T_{\mu\nu}$:
\[
\left\{
\begin{aligned}
  S_{00} &=\overset{(0)}{S_{00}}+\overset{(2)}{S_{00}}+\dots, \\
  S_{ij} &=\overset{(2)}{S_{ij}}+\overset{(4)}{S_{ij}}+\dots,
\end{aligned}
\right.
\]
where $\overset{(N)}{S_{\mu\nu}}\sim\left(M/\bar{r}^3\right)\bar{v}^N$.

The above immediately yields the required terms of the expansion of the right-hand side of the Einstein equations:
\begin{equation}
    \overset{(0)}{S_{00}}=\frac{2}{3}\overset{(0)}{T^{00}},\quad\overset{(0)}{S_{ij}}=\frac{1}{3}\delta_{ij}\overset{(0)}{T^{00}},\quad\overset{(2)}{S_{00}}=\frac{2}{3}\left[\overset{(2)}{T^{00}}-2\overset{(2)}{g_{00}}\overset{(0)}{T^{00}}+\frac{1}{2}\overset{(2)}{T^{mm}}\right].
    \label{eq:13}
\end{equation}
Equations \eqref{eq:13} differ by coefficients from similar ones in four-dimensional spacetime \cite{Weinberg:1972}, which prevents reducing the equation for $\overset{(4)}{g_{00}}$ to a linear one.

The obtained results allow us to write down the linearized Einstein equations:
\begin{equation}
\left\{
\begin{aligned}
  &-\frac{1}{2}\partial^2_m\overset{(2)}{g_{00}}=\frac{16}{3}\pi G_5\overset{(0)}{T^{00}},\\
  &\frac{1}{2}\left(\partial_m\partial_j\overset{(2)}{g_{mi}}+\partial_m\partial_i\overset{(2)}{g_{mj}}-\partial_m^2\overset{(2)}{g_{ij}}+\partial_j\partial_i\overset{(2)}{g_{00}}-\partial_j\partial_i\overset{(2)}{g_{mm}}\right)=\frac{8}{3}\pi G_5\delta_{ij}\overset{(0)}{T^{00}},
\end{aligned}
\right.
\label{eq:14}
\end{equation}
together with the equation for $\overset{(4)}{g_{00}}$:
 \[
 \frac{1}{2}\partial^2_m\overset{(4)}{g_{00}}=\partial_m\partial_0\overset{(3)}{g_{m0}}+\frac{1}{2}\partial_m\overset{(2)}{g_{mj}}\partial_j\overset{(2)}{g_{00}}+\frac{1}{2}\overset{(2)}{g_{mj}}\partial_m\partial_j\overset{(2)}{g_{00}}-\frac{1}{2}\partial_0^2\overset{(2)}{g_{mm}}-\frac{1}{4}\partial_k\overset{(2)}{g_{00}}\partial_k\overset{(2)}{g_{mm}}-
 \]
\begin{equation}
     -\left(\frac{1}{2}\partial_m\overset{(2)}{g_{00}}\right)^2-\frac{16}{3}\pi G_5\left[\overset{(2)}{T^{00}}-2\overset{(2)}{g_{00}}\overset{(0)}{T^{00}}+\frac{1}{2}\overset{(2)}{T^{mm}}\right].
     \label{eq:15}
\end{equation}

In the Newtonian limit, we use following spherical substitution for gravity field:
\[
    \overset{(2)}{g_{00}}=-2F,\quad\overset{(2)}{g_{ij}}=-\delta_{ij}F.
\]
Then the equations in \eqref{eq:14} are rewritten in a unified form:
\[
\partial^2_mF=\frac{16}{3}\pi G_5\overset{(0)}{T^{00}}.
\]
To reproduce the Poisson equation for the Newtonian gravitational potential, we redefine the five-dimensional gravitational constant
\[
G_5^*=\frac{8}{3\pi}G_5
\]
As a result, the linearized Einstein equations in the first order of approximation are derived
\begin{equation}
    \partial^2_mF=2\pi^2 G_5^*\overset{(0)}{T^{00}}.
    \label{eq:16}
\end{equation}
From \eqref{eq:16} one can restore the gravitational action in the Newtonian limit
\begin{equation}
    S_{NR,G}=\int d^5x\left[-\frac{1}{6\pi^2 G_5^*}(\partial_i F)^2\right].
    \label{eq:17}
\end{equation}
It should be noted that this result does not require the use of harmonic or other gauges.

\subsection{Non-relativistic limit for a complex scalar field} \label{NR complex field}

Construction of a simple non-relativistic limit for complex field requires the introduction of following field substitution in the form of \cite{Namjoo:2017}:

\begin{equation}
    \varphi\left(t,\vec{x}\right)=\frac{1}{\sqrt{2m}}e^{-imt}\psi\left(t,\vec{x}\right).
    \label{eq:18}
\end{equation}
It is worth noting at once that this work studies stationary solutions, i.e., the ansatz $\psi(t,\vec{x})=e^{-i\gamma t}\phi(r)$ is used for the simple non-relativistic limit and $\varphi(t,\vec{x})=(1/\sqrt{8\pi})e^{-i\omega t}\kappa(r)$ in the relativistic case (Sec. \ref{chapter 3}). This circumstance allows one to relate the small parameter $\bar{v}^2$, previously used in deriving the Newtonian approximation and (PN) corrections, to the small expansion parameter for the complex scalar field $|m^2-\omega^2|/m^2$. Under the condition relating the quantities $\omega$ and $\gamma$ ($\omega=m+\gamma$), we obtain the relation:
\[
\frac{m^2-\omega^2}{m^2}\approx\frac{2\gamma}{m}=\bar{v}^2.
\]
As a result, the non-relativistic limit for the scalar field and the Newtonian approximation are constructed in the same way using the expansion in $\bar{v}^2$. In what follows, we use this result to discard terms of the next order in the small parameter in the matter Lagrangian density $\mathcal{L}_M$.

$\mathcal{L}_M$ is given by:
\[
\mathcal{L}_M=\sqrt{-g}\left(-g^{00}\partial_0\varphi^*\partial_0\varphi-g^{ij}\partial_i\varphi^*\partial_j\varphi-m^2\varphi^*\varphi\right).
\]
The ansatz \eqref{eq:18} and the expansion for $g^{\mu\nu}$ and $\sqrt{-g}$ up to terms $\bar{v}^2$ inclusive are substituted into $\mathcal{L}_M$. It is also used that the derivative $\partial_0$ of $\psi$ multiplies the component by $\gamma/m$, which increases the order in the small parameter by $\bar{v}^2$, as discussed above. This circumstance can be expressed by the inequality: $|\partial_0\psi|^{2}\ll m|\partial_{0}\psi|\ll m^{2}|\psi|^{2}$.

Terms having the structure $\sim\left(F/2m\right)\partial_i\psi^*\partial_j\psi$, $\left|\partial_0\psi\right|^2/2m$ and $F^2\left|\psi\right|^2$ are discarded. This leads to the action in the non-relativistic limit without corrections of the form:
\[
    S_{NR}=\int d^5x\left[\frac{i}{2}\left(\psi^*\partial_0\psi-\psi\partial_0\psi^*\right)-\frac{|\partial_i\psi|^2}{2m}-\frac{3}{2}mF|\psi|^2-\frac{1}{6\pi^2 G_5^*}\left(\partial_i F\right)^2\right].
\]

The expression for $S_{NR}$ has a structure similar to that of the five-dimensional generalization of the Wick--Cutkosky model \cite{nugaev2017}, which, as will be shown below, is important for the analysis.


\subsection{Field equations and their solution} \label{Field equations and their solution}

For further analysis of the equations of motion and for ease of comparison with the five-dimensional generalization of the Wick--Cutkosky model, it is convenient to redefine the field $F$ as: $f=\sqrt{\frac{2m}{3\pi^2G^*_5}}F$. Then $S_{NR}$ takes the form:
\begin{equation}
    S_{NR}=\int d^5x\left[\frac{i}{2}(\psi^*\partial_0\psi-\psi\partial_0\psi^*)-\frac{|\partial_i\psi|^2}{2m}-b^2f|\psi|^2-\frac{(\partial_{i} f)^2}{4m}\right],
    \label{eq:19}
\end{equation}
where $b^2=\sqrt{\frac{27}{4}m\pi^2G_5^*}$.

Now it is useful to obtain the equations of motion and discuss the relation between the theory under consideration and the five-dimensional generalization of the Wick--Cutkosky model. Varying \eqref{eq:19} yields the following equations of motion:
\[
\left\{
\begin{aligned}
& i\partial_0\psi-b^2f\psi+\frac{\Delta\psi}{2m}=0, \\
& -b^2\psi^*\psi+\frac{\Delta f}{2m}=0.
\end{aligned}
\right.
\]
In this subsection, we consider spherically symmetric solution given by stationary ansatz: $\psi(t,\vec{x})=e^{-i\gamma t}\phi(r)$, $f=f(r)$, thus the radial part of the Laplacian operator is $\Delta = \partial_{r}^{2} + \frac{3}{r} \partial_{r}$.
\begin{align}
\left\{
\begin{aligned}
& -2m\gamma\phi-\Delta\phi+2mb^2f\phi=0, \\
& -\Delta f+2mb^2\phi^2=0,
\end{aligned}
\right.
\label{eq:20}
\end{align}
with boundary conditions: regularity at the origin and vanishing fields at infinity. We restrict our consideration to monotonic profile functions $f(r)$, i.e. we do not study radial excitations of BSs. 
\begin{align}
\left\{
\begin{aligned}
\partial_rf|_{r=0}=0\quad f|_{r \to \infty}=0, \\
\partial_r\phi|_{r=0}=0\quad\phi|_{r \to \infty}=0.
\end{aligned}
\right.
\label{eq:21}
\end{align}
The system \eqref{eq:20} is similar to the one that can be obtained in the article \cite{nugaev2017}. In this theory, a complex massless scalar field $\xi$ is considered, interacting through a Yukawa term with a massless real scalar field $\hat{\Psi}$. The action of this model is written as:
\[
S_{Wick-Cutkosky}=\int d^5x\left(\partial_\mu\xi^*\partial^\mu\xi-B^2\xi^*\xi +\frac{1}{2}\partial_\mu\tilde{\Psi}\partial^\mu\tilde{\Psi}-h\tilde{\Psi}\xi^*\xi\right),
\]
where $h$ is the Yukawa coupling constant, $B^2 = h\Psi_0$, $\Psi_0 \equiv const$ is the vacuum expectation value for the field $\hat{\Psi} = \Psi_0 + \tilde{\Psi}$.


Next, we obtain the equations of motion for $S_{\text{Wick--Cutkosky}}$:
\[
\left\{
\begin{aligned}
& \square\xi+h\hat{\Psi}\xi=0, \\
& \square\hat{\Psi}+h\xi^*\xi=0.
\end{aligned}
\right.
\]
Spherically symmetric stationary nodeless solutions for $\xi$ and $\tilde{\Psi}$ of the form $\xi=e^{-i\omega t}\Xi(r)$, $\tilde{\Psi}=\tilde{\Psi}(r)$ are also considered.
\begin{align}
\left\{
\begin{aligned}
& -\omega^2\Xi-\Delta \Xi+B^2\Xi+h\tilde{\Psi}\Xi=0, \\
& -\Delta\tilde{\Psi}+h\Xi^2=0,
\end{aligned}
\right.
\label{eq:22}
\end{align}
where $B^2=h\Psi_0\geq0$.

For the case $\Psi_0=0$ equations \eqref{eq:20} and \eqref{eq:22} are identical.

Next we numerically find the solution of the system \eqref{eq:20} with the boundary conditions \eqref{eq:21}. For convenience of calculations, a reparametrization procedure is performed.
\begin{equation}
    X=r\sqrt{-2m\gamma}\quad \mathfrak{G}(X)=\frac{2mb^2}{-2m\gamma}f(r)\quad \Phi(X)=\frac{2mb^2}{-2m\gamma}\phi(r).
    \label{eq:r1}
\end{equation}
The quantities $X$, $\mathfrak{G}(X)$, $\Phi(X)$ are dimensionless. They bring \eqref{eq:20} and \eqref{eq:21} to the form:
\[
\left\{
\begin{aligned}
& -\Delta_X\Phi+\Phi+\mathfrak{G}\Phi=0, \\
& -\Delta_X\mathfrak{G}+\Phi^2=0,
\end{aligned}
\right.
\]
with boundary conditions:
\[
\left\{
\begin{aligned}
\partial_X\mathfrak{G}|_{X=0}=0,\quad\mathfrak{G}|_{X\rightarrow\infty}=0, \\
\partial_X\Phi|_{X=0}=0,\quad\Phi|_{X\rightarrow\infty}=0.
\end{aligned}
\right.
\]
For subsequent use of the shooting method, it is convenient to make one more substitution \cite{nugaev2017}.
\begin{equation}
\Phi\left(X\right)=\left(C-1\right)\hat{\Phi}\left(Y\right),\quad\mathfrak{G}\left(X\right)=-C+\left(C-1\right)\hat{G}\left(Y\right),\quad Y=\sqrt{C-1}X,
\label{eq:r2}
\end{equation}
where: $C=-\mathfrak{G}(X=0)$.
\begin{align}
\left\{
\begin{aligned}
& -\Delta_Y\hat{\Phi}-\hat{\Phi}+\hat{\Phi}\hat{G}=0, \\
& -\Delta_Y\hat{G}+\hat{\Phi}^2=0.
\end{aligned}
\label{eq:1.20}
\right.
\end{align}
As a result, \eqref{eq:r2} gives the condition for $\mathfrak{G}|_{Y\rightarrow\infty}=C/(C-1)$, while for the integration of \eqref{eq:1.20} the following conditions are applied:
\begin{align}
\left\{
\begin{aligned}
& \partial_Y\hat{G}|_{Y=0}=0,\quad\hat{G}|_{Y=0}=0, \\
& \partial_Y\hat{\Phi}|_{Y=0}=0,\quad\hat{\Phi}|_{Y\rightarrow\infty}=0.
\end{aligned}
\label{eq:1.21}
\right.
\end{align}
For the system \eqref{eq:1.20} with boundary conditions \eqref{eq:1.21}, a numerical shooting method is implemented, in which the value of the field $\hat{\Phi}(Y=0)$ is iteratively adjusted to satisfy the condition $\hat{\Phi}_{Y\rightarrow\infty}=0$. At each step, the system of equations is solved using the fourth-order Runge--Kutta method on a uniform grid. As a result, the following results are obtained:
\begin{figure}[htbp]
    \centering
    \includegraphics[width=\textwidth]{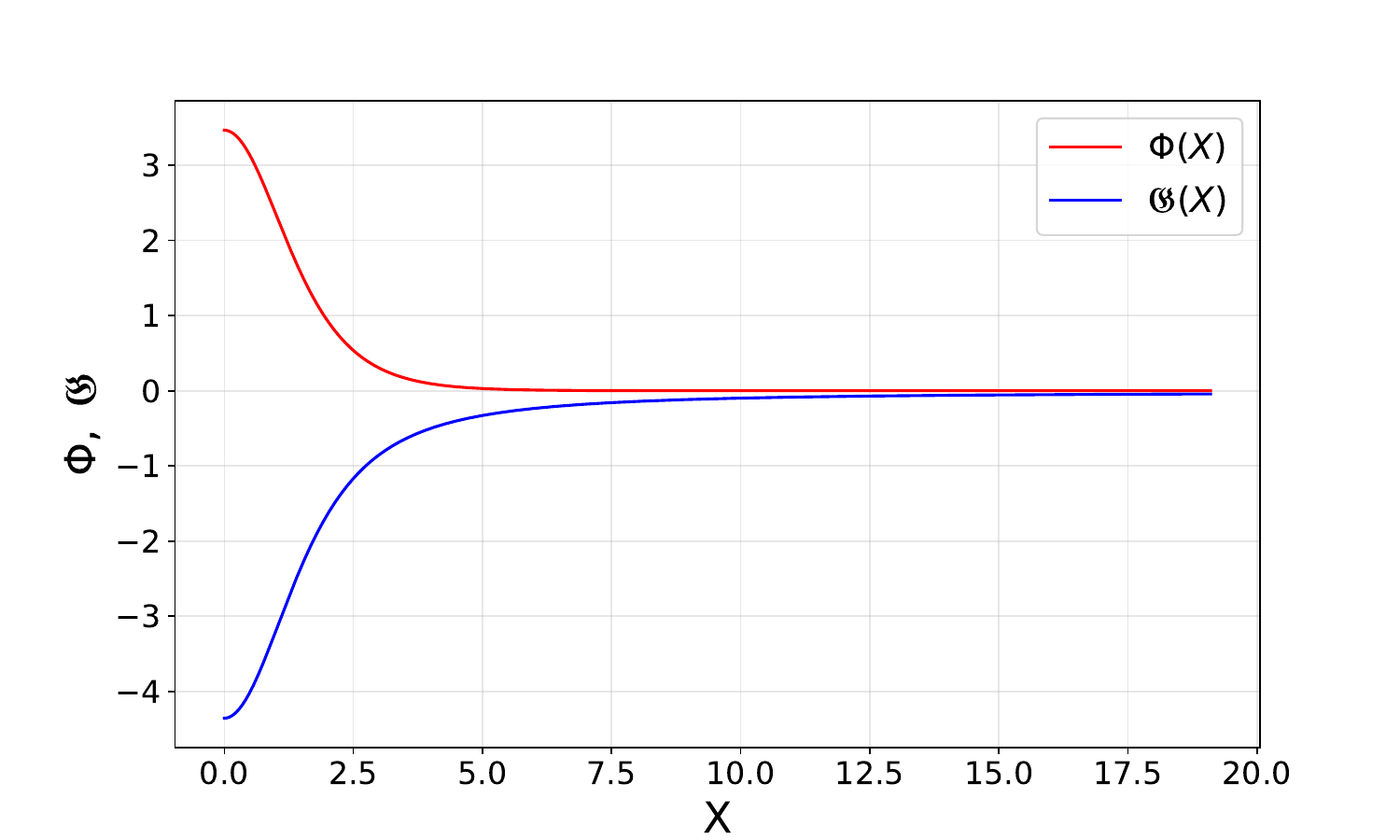}
    \caption{Profiles of the fields $\Phi(X)$ (red solid line) and $\mathfrak{G}(X)$ (blue solid line).}
    \label{Fig:2}
\end{figure}
\begin{figure}[htbp]
    \centering
    \includegraphics[width=1.0\textwidth]{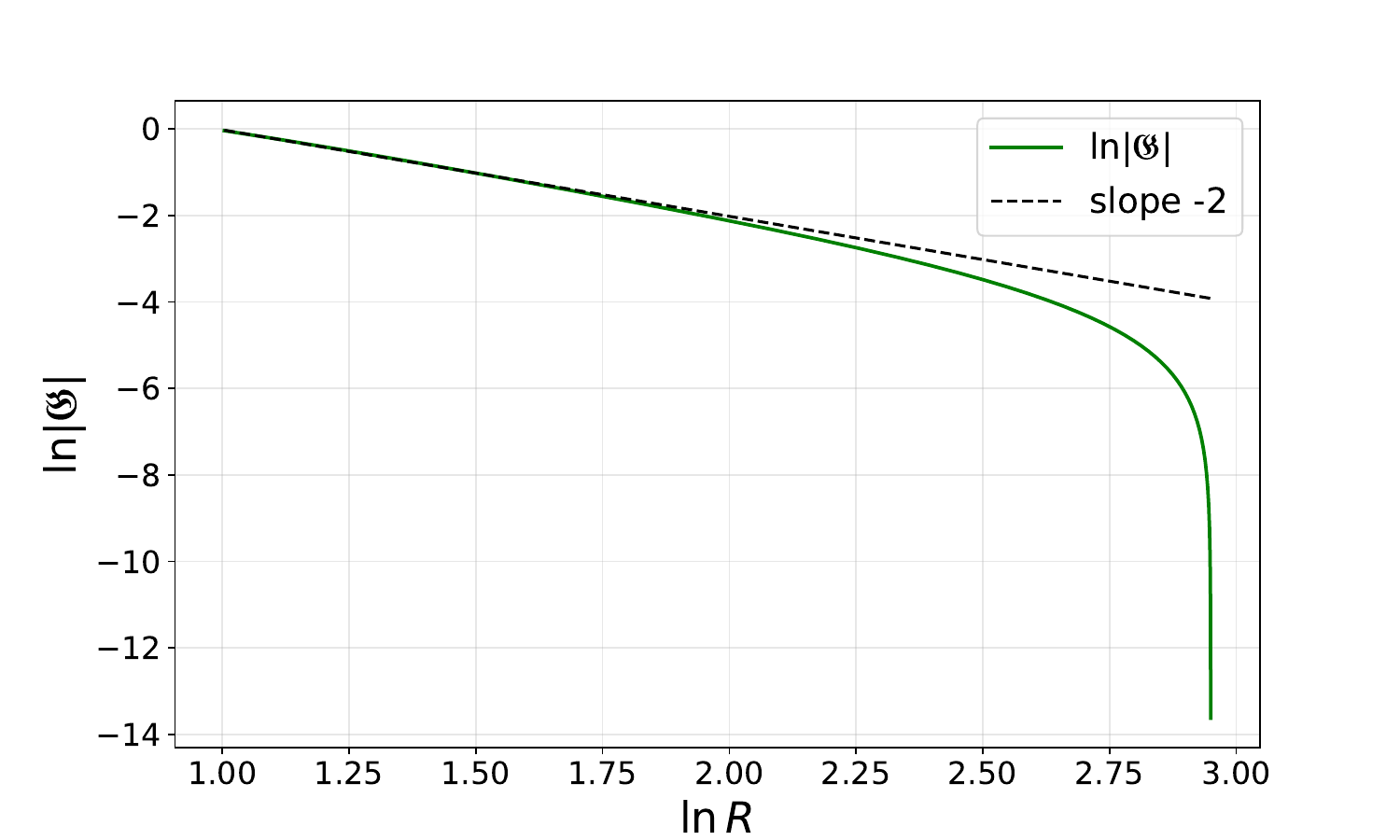}
    \caption{Profile of the field $\mathfrak{G}(X)$ (green solid line). The black dashed line corresponds to the $1/X^2$ dependence.}
    \label{Fig:3}
\end{figure}

Fig. \ref{Fig:2} shows the profiles of the fields, whose absolute values decrease to zero as $X\rightarrow\infty$. At the same time, $\mathfrak{G}(X)$ exhibits quadratic decay (Fig. \ref{Fig:3}), which is consistent with the transition of the equation of motion for this field into the Laplace equation (see \eqref{eq:1.20}) when $\Phi(X)|_{X\rightarrow\infty}\rightarrow\infty$.

From Figs. \ref{Fig:2} and \ref{Fig:3} it follows that the solution found is localized. When all relativistic corrections for the complex scalar field are taken into account, the theory described by the action \eqref{eq:19} goes over into the five-dimensional generalization of the Wick--Cutkosky model. In this model, a similar class of localized solutions is represented by Q-balls. For them, there exists a stable branch, as well as an unstable one, located between the scale-invariance point and a classically unstable solution with zero $U(1)$ charge and nonzero energy, commonly called a sphaleron. Therefore, the inclusion of relativistic corrections can lead to either a stable or an unstable field configuration.


\section{Physical properties}\label{chapter 2}

\subsection{Energy and $U(1)$ charge}
We examine the properties of the solution found in Sec. \ref{Field equations and their solution}, using the relation between the $U(1)$ charge (see more details in App. \ref{Integral characteristics}) and the non-relativistic energy H.

The Legendre transformation is used to find the Hamiltonian of the system:
\[
\text{H}=\int d^4x\Bigl(\pi_{\psi}\partial_0{\psi}+\pi_{\psi^*}\partial_0{\psi^*}-\mathcal{L}\Bigr);\qquad\pi_{\psi}=\frac{\partial\mathcal{L}}{\partial(\partial_0{\psi})}=i\psi^*,\qquad\pi_{\psi^*}=0,
\]
where $\mathcal{L}$ is the subintegral expression of action \eqref{eq:19}.

Upon substitution of the quantities and integration over the angular variables we obtain:
\[
    \text{H}=\frac{\pi^2}{2m}\int\left(2(\nabla\phi)^2+4mb^2f\phi^2+(\nabla f)^2\right)r^3 dr.
\]
Using the equations of motion, one can express $(\nabla\phi)^2$ and establish a relation between H and $N$:
\[
\text{H}=\frac{\pi^2}{2m}\int \left(2\phi(-\Delta\phi+2mb^2f\phi)+(\nabla f)^2\right)r^3dr=\frac{\pi^2}{2m}\int \left(4m\gamma\phi^2+(\nabla f)^2\right)r^3dr.
\]
The first component in the last equality is the $U(1)$ charge $N$ multiplied by $\gamma$.
\begin{equation}
    \text{H}=\gamma N+\frac{\pi^2}{2m}\int(\nabla f)^2r^3dr>\gamma N
    \label{eq:2.2}
\end{equation}
Another relation between H and $N$ can also be obtained. It allow us to understand the meaning of the quantity $\gamma$. For this purpose, derivatives of H and $N$ with respect to $\gamma$ are taken using the equations of motion \eqref{eq:20}. As a result:
\[
\frac{d\text{H}}{d\gamma}=4\pi^2\gamma\int r^3\phi\frac{d\phi}{d\gamma}dr,\quad\frac{dN}{d\gamma}=4\pi^2\int r^3\phi\frac{d\phi}{d\gamma}dr.
\]
Then:
\begin{equation}
    \frac{d\text{H}}{d\gamma}=\gamma\frac{dN}{d\gamma}.
    \label{eq:2.3}
\end{equation}
Relation \eqref{eq:2.3} is a well-known result for Q-balls. The quantity $\gamma$ can accordingly be interpreted as the non-relativistic chemical potential of the system.

For further numerical analysis, the quantities $N$ and H are expressed in terms of dimensionless variables. After direct substitution \eqref{eq:r1} into \eqref{eq:2.1} and \eqref{eq:2.2}:
\begin{equation}
    N=\frac{I}{(2mb^2)^2};\qquad I=2\pi^2\int R^3\Phi^2dR.
    \label{eq:2.4}
\end{equation}

\begin{equation}
    \text{H}=\gamma N-\frac{\gamma}{(2mb^2)^2}J;\qquad J=\pi^2\int R^3(\partial_R\mathfrak{G})^2dR.
    \label{eq:2.5}
\end{equation}
Using equation \eqref{eq:2.3}, the relation between $J$ and $I$ can be established.
\[
\frac{d\text{H}}{d\gamma}=N+\gamma\frac{dN}{d\gamma}-\frac{1}{(2mb^2)^2}J\Rightarrow
J=I
\]
From the equality above and formula \eqref{eq:2.5}, one can obtain an important result:
\begin{equation}
    \text{H}=0.
    \label{eq:2.6}
\end{equation}
For the convenience of calculations, the value of $I$ are expressed using substitution \eqref{eq:r2}:
\begin{equation}
    I=2\pi^2\int Y^3\Phi^2dY;\qquad C=\frac{\hat{G}(\infty)}{\hat{G}(\infty)-1}.
    \label{eq:2.7}
\end{equation}
According to the result of numerical integration: $I=303.78\quad J=87.62\quad C=4.51$.

From \eqref{eq:2.6} and \eqref{eq:2.7}, it follows that the quantities $N$ and H are independent of $\gamma$, a result stemming from the scale invariance of the theory under consideration and the conservation of the corresponding dilatation current. This observation also corresponds to energy degeneracy at the single point $\text{H}=0$ (see more details in the next subsection).
The charge $N$ is the same for all non-relativistic BSs. The well-known relation $d\text{H}/dN=\gamma$ from the analysis of Q-balls does not hold; instead, only its analog \eqref{eq:2.3} is valid, since $d\text{H}/d\gamma=0$ from \eqref{eq:2.6}. Both of these properties are also a consequence of the scale invariance \eqref{eq:19}. Below we will consider the restoration of scale invariance in the non-relativistic limit.

\subsection{Scale invariance and power counting}

A peculiarity of the non-relativistic limit is the restoration of scale invariance \cite{Duval:2015gba, Duval:2024eod} in \eqref{eq:19} for the originally non-symmetric relativistic action \eqref{eq:1}. Scale symmetry, as will be presented in this subsection, is the reason that the non-relativistic energy of the previously found solution (see \eqref{eq:2.6}) does not depend on the scale of the theory and is equal to zero at a nonzero $U(1)$ charge. Such a solution is called critical.

Before proceeding to the discussion of the consequences of scale invariance, it is worth showing the invariance of the action \eqref{eq:19} under the transformations of this group. The action of the group on the fields is defined as:
\begin{equation} \psi\rightarrow\sigma^\alpha\psi(\sigma^2t,\sigma\vec{x}),\qquad F\rightarrow\sigma^\beta F(\sigma^2t,\sigma\vec{x}),
\label{eq:2.8}
\end{equation}
the action takes the form
\begin{equation}
    S_{NR}=\int d^5x\Bigl[\sigma^{2\alpha+2-6}i\dot{\psi}\psi^*-\sigma^{2\alpha+\beta-6}b^2f\psi\psi^*-\sigma^{2\alpha+2-6}\frac{|\nabla\psi|^2}{2m}-\sigma^{2\beta+2-6}\frac{(\nabla f)^2}{4m}\Bigr].
    \label{eq:2.9}
\end{equation}
From \eqref{eq:2.9} it follows that $S_{NR}$ is invariant under transformation \eqref{eq:2.8}
\begin{equation}
    \alpha=2,\qquad \beta=2.
    \label{eq:2.10}
\end{equation}
Applying Noether's method for such transformations yields the corresponding conserved current:
\begin{equation}
    J^D_0=\frac{ix_j}{2}(\psi^*\partial_j\psi-\psi\partial_j\psi^*)+\frac{x_0}{m}|\partial_j\psi|^2+2b^2x_0f|\psi|^2+\frac{x_0}{2m}(\partial_jf)^2,
\end{equation}
\[
    J^D_j=-\frac{1}{m}(\psi^*\partial_j\psi+\psi\partial_j\psi^*)-\frac{x_0}{m}(\partial_0\psi^*\partial_j\psi+\partial_0\psi\partial_j\psi^*)-
\]
\[
-\frac{x_i}{2m}(\partial_j\psi^*\partial_i\psi+\partial_j\psi\partial_i\psi^*)-\frac{1}{m}f\partial_jf-\frac{x_0}{m}\partial_jf\partial_0f-\frac{x_i}{2m}\partial_jf\partial_if-
\]
\begin{equation}
    -\frac{ix_j}{2}(\psi^*\partial_0\psi-\psi\partial_0\psi^*)+\frac{x_j}{2m}|\partial_i\psi|^2+b^2x_jf|\psi|^2+\frac{x_j}{4m}(\partial_if)^2.
\end{equation}
As noted earlier, the criticality of the solution found is a consequence of the scale invariance of the system. The result $\text{H}=0$ can be obtained from the conservation of the dilatation current. Together with the decay of the fields at infinity, this leads to:
\[
\int d^4x\left(\partial_0J^D_0\right)=0.
\]
Taking into account the ansatz \eqref{eq:21} and the form of the solution sought:
\[
\int d^4x\partial_0\left(\frac{1}{m}[\nabla\phi(r)]^2+2b^2f(r)[\phi(r)]^2+\frac{1}{2m}[\nabla f(r)]^2\right)=0=2m\text{H}.
\]
Thus, $\text{H}=0$ is indeed a consequence of the scale invariance of the system. The introduction of corrections of the next order in the small parameter into the theory will cause a weak violation of scale symmetry and remove the degeneracy in H.

The method described in \cite{burgess2021} allows one to determine the orders of the contributions from the operators in \eqref{eq:19}.
The action \eqref{eq:19} must be dimensionless in the natural system of units $\hbar=c=1$, which implies that components with dimensionality higher than $[m^5]$ are suppressed by a large mass scale in the low‑energy limit. The single small expansion parameter $\bar{v}^2$ ensures that terms associated with both the gravitational field and the complex scalar field are suppressed in the same manner.

According to \cite{burgess2021}, the relevance of the components of \eqref{eq:19} can be determined by their transformation properties under scale transformations. From equations \eqref{eq:2.8} - \eqref{eq:2.10} we have:
\begin{equation}
\left\{
\begin{aligned}
& \partial_0\sim\sigma^2,\quad\partial_i\sim\sigma,\\
& \psi\sim f\sim\sigma^2.
\end{aligned}
\right.
\label{eq:2.13}
\end{equation}
In fact, $\sigma$ correctly counts the power of $\bar{v}$, taking into account that derivatives also contribute a factor of $1/\bar{r}$.

Using the above analysis, one can establish that all terms of the action \eqref{eq:19} are terms of order $\sigma^6$ (see Fig. \ref{Fig:2}). It also becomes clear that any inclusion of next‑order corrections violate the scale invariance.

\section{Weak violation of scale invariance  and kinematical instability of 5D Bosоn star}\label{chapter 3}

In the previous sections, the simple non-relativistic limit of the action \eqref{eq:1} was considered. For this limit, the restoration of scale invariance was found and a critical boson star type solution was obtained. In this section, we investigate the weak violation of scale invariance by introducing higher-order corrections and also consider the problem of the kinematic instability of field configurations. Corrections for a complex scalar field associated with antiparticle degrees of freedom could be introduced. Indeed, one can consider following simple ansatz 
\[
\varphi=\frac{1}{\sqrt{2m}}(e^{-imt}\tilde{\zeta}+e^{imt}\tilde{\chi}^*),
\]
so that high-frequency degrees of freedom could be integrated-out in a time-dependent effective theory \cite{Eby:2018ufi}. 

In this section, we make use of a different approach to non-relativistic limit for the initial relativistic action for scalar field
\begin{equation}
    S_M=\int d^5x\sqrt{-g}\Bigl(-g^{00}\partial_0\varphi^*\partial_0\varphi-g^{ij}\partial_i\varphi^*\partial_j\varphi-m^2|\varphi|^2\Bigr),
    \label{eq:3.1}
\end{equation}
which preserve global $U(1)$ symmetry from high-energy theory. We define fields $\zeta$ and $\chi$ \cite{Belova:1989qb, Bjorken:1964rqm} as
\begin{equation}
    \zeta=\frac{1}{2}\left(\varphi+\frac{i}{m}\partial_0\varphi\right),\quad\chi=\frac{1}{2}\left(\varphi-\frac{i}{m}\partial_0\varphi\right)\Rightarrow\varphi=\zeta+\chi,\quad\partial_0\varphi=-im(\zeta-\chi).
    \label{eq:3.2}
\end{equation}
Next, the matter action \eqref{eq:3.1} is rewritten using \eqref{eq:3.2}
\[
S_M=\int d^5x\Bigl[-g^{00}\Bigl(im\left[\zeta^*\partial_0\zeta-\zeta\partial_0\zeta^*+\chi\partial_0\chi^*-\chi^*\partial_0\chi\right]+
\]
\[
+im\left[\zeta^*\partial_0\chi-\zeta\partial_0\chi^*+\chi\partial_0\zeta^*-\chi^*\partial_0\zeta\right]\Bigr)-g^{ij}\partial_i\left(\zeta^*+\chi^*\right)\partial_j\left(\zeta+\chi\right)-
\]
\[
-m^2\left(1+g^{00}\right)(|\zeta|^2+|\chi|^2)-m^2\left(1-g^{00}\right)\left(\zeta^*\chi+\chi^*\zeta\right)\Bigr].
\]
The non-relativistic limit for the fields $\zeta$ and $\chi$ can be taken by substitution: $(\zeta, \chi) \rightarrow \frac{e^{-imt}}{\sqrt{2m}}(\zeta, \chi)$, where we emphasize that the same prefactor is singled-out for both fields in accordance with initial global $U(1)$ symmetry.  

Then the $S_M$ is rewritten as:
\[
    S_M=\int d^5x \sqrt{-g}\Bigl(-ig^{00}(\zeta^*\partial_0{\zeta}-\chi^*\partial_0{\chi})-\frac{g^{ij}}{2m}\partial_i(\zeta^*+\chi^*)\partial_j(\zeta+\chi)-
\]
\begin{equation}
    -\frac{m}{2}(g^{00}+1)[|\zeta|^2+\chi^*\zeta+\zeta^*\chi]+\frac{m}{2}(3g^{00}-1)|\chi|^2\Bigr).
    \label{eq:3.3}
\end{equation}
Let us check the validity of this approach. We immediately note that the action \eqref{eq:3.3} is $U(1)$ invariant, which was observed in the previous section for \eqref{eq:19} and must hold in the relativistic theory. The symmetry is preserved. It can also be shown how the field $\chi$ is suppressed and the correct number of degrees of freedom is preserved. Variation of the action \eqref{eq:3.3} with respect to the fields $\zeta$ and $\chi$ leads to the following field equations:
\[
\left\{
\begin{aligned}
& -i\sqrt{-g}g^{00}\partial_0{\zeta}=\frac{m}{2}(g^{00}+1)(\zeta+\chi)-\frac{g^{ij}}{2m}\partial_i\partial_j(\zeta+\chi)-\frac{1}{2m}\partial_ig^{ij}\partial_j(\zeta+\chi), \\
& i\sqrt{-g}g^{00}\partial_0{\chi}=\frac{m}{2}(g^{00}+1)\zeta+\frac{m}{2}(3g^{00}-1)\chi-\frac{g^{ij}}{2m}\partial_i\partial_j(\zeta+\chi)-\frac{1}{2m}\partial_ig^{ij}\partial_j(\zeta+\chi),
\end{aligned}
\right.
\]
The constraint relation connecting $\zeta$ and $\chi$ follows from this system:
\begin{equation}
    i\sqrt{-g}g^{00}(\partial_0{\zeta}+\partial_0{\chi})=m(g^{00}-1)\chi.
    \label{eq:3.4}
\end{equation}
In the non‑relativistic limit, the formal solution can be expanded in a series under consideration
\begin{equation}
    \chi=\frac{i\sqrt{-g}g^{00}\partial_0\zeta}{m(g^{00}-1-(ig^{00}\sqrt{-g}/m)\partial_0)}=0-\frac{i}{m}g^{00}\sqrt{-g}\partial_0\zeta+\dots
    \label{eq:3.5}
\end{equation}
From \eqref{eq:3.5} it follows that the field $\chi$ is suppressed by mass scale, thus representing a small component of the relativistic field \cite{Bjorken:1964rqm}. Moreover, \eqref{eq:3.4} and its complex conjugate relation reduce the number of independent degrees of freedom from four to two, which corresponds to the two degrees of freedom of $\varphi$ in the relativistic theory.

For completeness, it is necessary to introduce post-Newtonian corrections. To do this, it is important to correctly take into account the expansion terms of $g_{\mu\nu}$. The form of these terms found earlier in Sec. \ref{chapter 1} is used.

We consider all components of \eqref{eq:3.3}.

The first term is:
\[
-\frac{i}{2}\sqrt{-g}g^{00}\left(\zeta^*\partial_0{\zeta}-\zeta\partial_0\zeta^*+\chi\partial_0\chi^*-\chi^*\partial_0{\chi}\right)=
\]
\begin{equation}
    =-\frac{i}{2}\left(\zeta^*\partial_0{\zeta}-\zeta\partial_0\zeta^*+\chi\partial_0\chi^*-\chi^*\partial_0{\chi}\right)\left(-1+3F-\frac{9}{2}F^2-\frac{1}{2}\left(\overset{(4)}{g_{00}}+\overset{(4)}{g_{ii}}\right)\right).
\end{equation}
The second term is:
\[
-\frac{\sqrt{-g}}{2m}g^{ij}\partial_i(\zeta^*+\chi^*)\partial_j(\zeta+\chi)=
\]
\begin{equation}
    =-\frac{1}{2m}\left(|\partial_i\zeta|^2+\partial_i\zeta^*\partial_i\chi+\partial_i\chi^*\partial_i\zeta+|\partial_i\chi|^2\right)\left(1-\frac{3}{2}F^2-\frac{1}{2}\left(\overset{(4)}{g_{00}}-\overset{(4)}{g_{ii}}\right)\right).
\end{equation}
The third term is:
\begin{equation}
    -\frac{m}{2}\sqrt{-g}\left(g^{00}+1\right)\left(|\zeta|^2+\chi^*\zeta+\zeta^*\chi\right)=-\frac{m}{2}\left(2F-6F^2-\overset{(4)}{g_{00}}\right)\left(|\zeta|^2+\chi^*\zeta+\zeta^*\chi\right).
\end{equation}
The fourth term is:
\begin{equation}
    \frac{m}{2}\sqrt{-g}\left(3g^{00}-1\right)|\chi|^2=\frac{m}{2}\left(-4+4F-21F^2-4\overset{(4)}{g_{00}}+\overset{(4)}{g_{ii}}\right)|\chi|^2.
\end{equation}
The result \eqref{eq:2.13}, equations \eqref{eq:15}, \eqref{eq:3.5}, and the fact that the field $\psi$ (in the simple non-relativistic limit of Sec. \ref{chapter 1}) and $\zeta$ describe particles, and hence are equivalent, allow us to write:
\begin{equation}
    \left\{
\begin{aligned}
& \partial_0\sim\sigma^2,\quad\nabla\sim\sigma,\\
& \zeta\sim\sigma^2,\quad\partial_0{\zeta}\sim\sigma^4,\quad\nabla\zeta\sim\sigma^3, \\
& F\sim\sigma^2,\quad\nabla F\sim\sigma^3,\\
& \chi\sim\sigma^4,\quad\partial_0{\chi}\sim\sigma^6,\quad\nabla\chi\sim\sigma^5, \\
& \overset{(4)}{g_{00}}\sim\sigma^4,\quad\overset{(4)}{g_{ij}}\sim\sigma^4.
\end{aligned}
\right.
\label{eq:3.10}
\end{equation}

In the complete action \eqref{eq:1} with corrections taken into account up to terms $\sigma^8$, one can separate out the components corresponding to the simple non-relativistic action and additional terms that weakly violate the scale symmetry of the system.
\[
S_{ext}=S_6+S_8,
\]
where
\begin{equation}
    S_{6}=\int d^5x\Bigl(\frac{i}{2}[\zeta^*\partial_0\zeta-\zeta\partial_0\zeta^*]-\frac{|\nabla\zeta|^2}{2m}-mF|\zeta|^2-\frac{1}{6\pi^2G_5^*}(\nabla F)^2\Bigr),
\end{equation}
\[
    S_{8}=\int d^5x\Bigl(-3iF(\zeta^*\partial_0\zeta-\zeta\partial_0\zeta^*)-\frac{\nabla\zeta^*\nabla\chi+\nabla\zeta\nabla\chi^*}{2m}+
\]
\begin{equation}
    +m\Bigl[3F^2+\frac{\overset{(4)}{g_{00}}}{2}\Bigr]|\zeta|^2-mF(\zeta^*\chi+\chi^*\zeta)-2m|\chi|^2-\frac{1}{12\pi^2G_5^*}(\nabla\overset{(4)}{g_{00}})^2\Bigr)
    \label{eq:3.12}
\end{equation}
The last component \eqref{eq:3.12} is effectively introduced to reproduce the right‑hand side of \eqref{eq:15}. $S_8$ contains terms $\sigma^8$ in the Lagrangian that are of lower order of contribution, $\bar{v}^8$, and cause a weak violation of scale invariance.

We have discussed the systematic inclusion of corrections of higher order in the small parameter, as well as how they cause a weak violating of scale symmetry. For the stability analysis, it is more convenient to start not from the full system of equations corresponding to \eqref{eq:1}, but from the relativistic theory. This issue was addressed in works
\cite{Brihaye:2015jja, Brihaye:2013zha, Hartmann:2010pm, franzin2024}. We will follow the approach outlined in \cite{franzin2024}. 
Hereafter, for the sake of convenience $M_{5Pl}=1$ is set. 
In that paper, the line element is given as:
\[
    ds^2=-e^{\nu(r)}dt^2+e^{\lambda(r)}dr^2+r^2d\Omega^2_3,
\]
where
\[
    \mathrm{d}\Omega_{D-2}^2 = \sum_{i=1}^{D-2} \Pi(i) \, \mathrm{d}\theta_i^2, \quad \Pi(i) = \begin{cases} 
1, & i = 1, \\ 
\prod_{j=1}^{i-1} \sin^2 \theta_j, & i > 1.
\end{cases}
\]
The functions $\nu(r)$ and $\lambda(r)$ are real metric functions.

The action is \eqref{eq:1}. For the complex scalar field we choose the harmonic ansatz, which is the relativistic generalization of \eqref{eq:18}:
\[
    \varphi(t,r)=\frac{1}{\sqrt{8\pi}}\kappa(r)e^{-i\omega t}.
\]
With this choice of ansatz, the system of differential equations for the metric functions and the scalar field is written as\footnote{The symbol $'$ in systems \eqref{eq:3.13} and \eqref{eq:3.14} denotes the derivative with respect to $r$.}:
\begin{equation}
\left\{
\begin{aligned}
& \lambda^{'}=\frac{1}{3}r(\kappa^{'})^{2}+\frac{1}{3}e^\lambda(m^2+e^{-\nu}\omega^2)\kappa^2-\frac{3}{r}(e^\lambda-1), \\
& \nu^{'}=\frac{1}{3}r(\kappa^{'})^{2}-\frac{1}{3}e^\lambda(m^2-e^{-\nu}\omega^2)\kappa^2+\frac{3}{r}(e^\lambda-1), \\
& \kappa^{''}=\Bigl(\frac{\lambda^{'}-\nu^{'}}{2}-\frac{3}{r}\Bigr)\kappa^{'}+e^\lambda(m^2-e^{-\nu}\omega^2)\kappa.
\end{aligned}
\right.
\label{eq:3.13}
\end{equation}
When imposing the boundary conditions, we require regularity at the origin.
\begin{equation}
        \left\{
\begin{aligned}
& \nu(0)=\nu_c,\quad\nu^{'}(0)=0,\quad\nu^{''}(0)=(3e^{-\nu_c}\omega^2-m^2)(\kappa_c^2/12), \\
& \lambda(0)=0,\quad\lambda^{'}(0)=0,\quad\lambda^{''}(0)=(e^{-\nu_c}\omega^2+m^2)(\kappa_c^2/12), \\
& \kappa(0)=\kappa_c,\quad\kappa^{'}(0)=0,\quad\kappa^{''}(0)=(m^2-e^{-\nu_c}\omega^2)(\kappa_c/8).
\end{aligned}
\right.
\label{eq:3.14}
\end{equation}
The metric functions $\nu(r)$ and $\lambda(r)$ must satisfy the condition of asymptotic flatness. The value $\nu(r=0)=\nu_c$ is arbitrary and in practice is normalized so that $e^{\nu(\infty)}=1$. Choosing a specific value of $\kappa_c$ turns the system \eqref{eq:3.13} into an eigenvalue problem for $\omega$. In the numerical implementation, the dimensionless variables are introduced by the rescaling $r\rightarrow m r$ and $\omega\rightarrow m\omega$, which is equivalent to setting $m=1$. The system \eqref{eq:3.13} is solved numerically by the shooting method, analogously to the original work \cite{franzin2024}.

Let us discuss the connection between the relativistic approach and the non-relativistic limit. We begin with a scalar field $\kappa$. For this field, the non-relativistic regime is realized as $\omega \to m$.
Therefore, the solution of system \eqref{eq:3.13} with $m=1$ should correspond to $\omega\rightarrow 1$ or $\bar{v}^2\rightarrow0$. The weak‑field limit of the gravitational field means that the metric functions deviate only slightly from zero ($|\nu|\ll 1$, $|\lambda|\ll 1$, or $\bar{v}^2\ll1$). Under these conditions, the solution of the relativistic system \eqref{eq:3.13} will correspond to the non‑relativistic limit with all higher‑order corrections taken into account.

Before presenting the numerical results, it is useful to examine the importance of the contribution of higher‑order gravitational corrections. This can be done by comparing the solution of system \eqref{eq:3.13} with those obtained in the relativistic Wick-Cutkosky model, which, by analogy, corresponds to the non‑relativistic limit but with higher‑order corrections taken into account only for the complex scalar field.

To carry out the analysis described above and to address the question of stability, we plot the total mass of the configuration $M_5$ versus the Noether charge $N_5$ associated with the $U(1)$ symmetry of the action \eqref{eq:1}. These quantities are defined as \cite{franzin2024} (see App. {Integral characteristics}).

Similarly, we plot the dependence of the configuration energy $E$ on the $U(1)$ charge $Q$ in the Wick--Cutkosky model, using their definitions according to Ref. \cite{nugaev2017} with the only modification for (4+1) dimensions. The resulting dependences are shown in Figs. \ref{Fig:4} and \ref{Fig:5}\footnote{The dashed and solid lines in them correspond to the characteristics found by solving the system of equations corresponding to the five-dimensional generalization of the Wick--Cutkosky model \cite{nugaev2017}, while the individual points correspond to the characteristics obtained by solving the system of equations \eqref{eq:3.13} with boundary conditions \eqref{eq:3.14}.}. 
\begin{center}
    \includegraphics[width=1\textwidth, trim=0 0 0 2.2cm, clip]{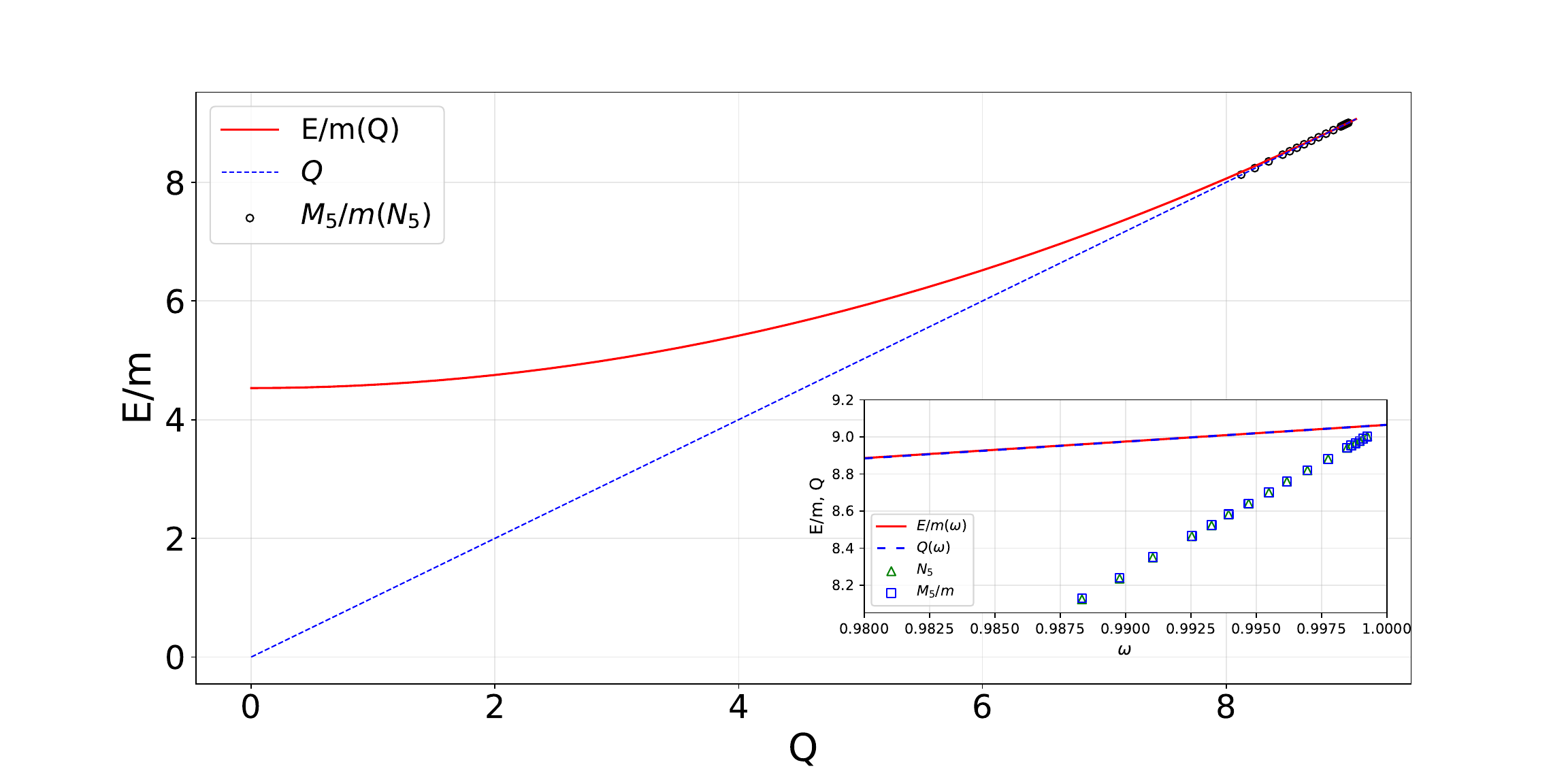}
    \captionof{figure}{Plot of the configuration energy versus the $U(1)$ charge for GR and the Wick--Cutkosky model. The inset shows the dependence of the energy (red solid line) and the $U(1)$ charge (blue dashed line) on the frequency $\omega$. The triangles and squares denote the points $N_5(\omega)$ and $M_5(\omega)$ in GR, respectively.}
    \label{Fig:4}
\end{center}
\begin{center}
    \includegraphics[width=\textwidth]{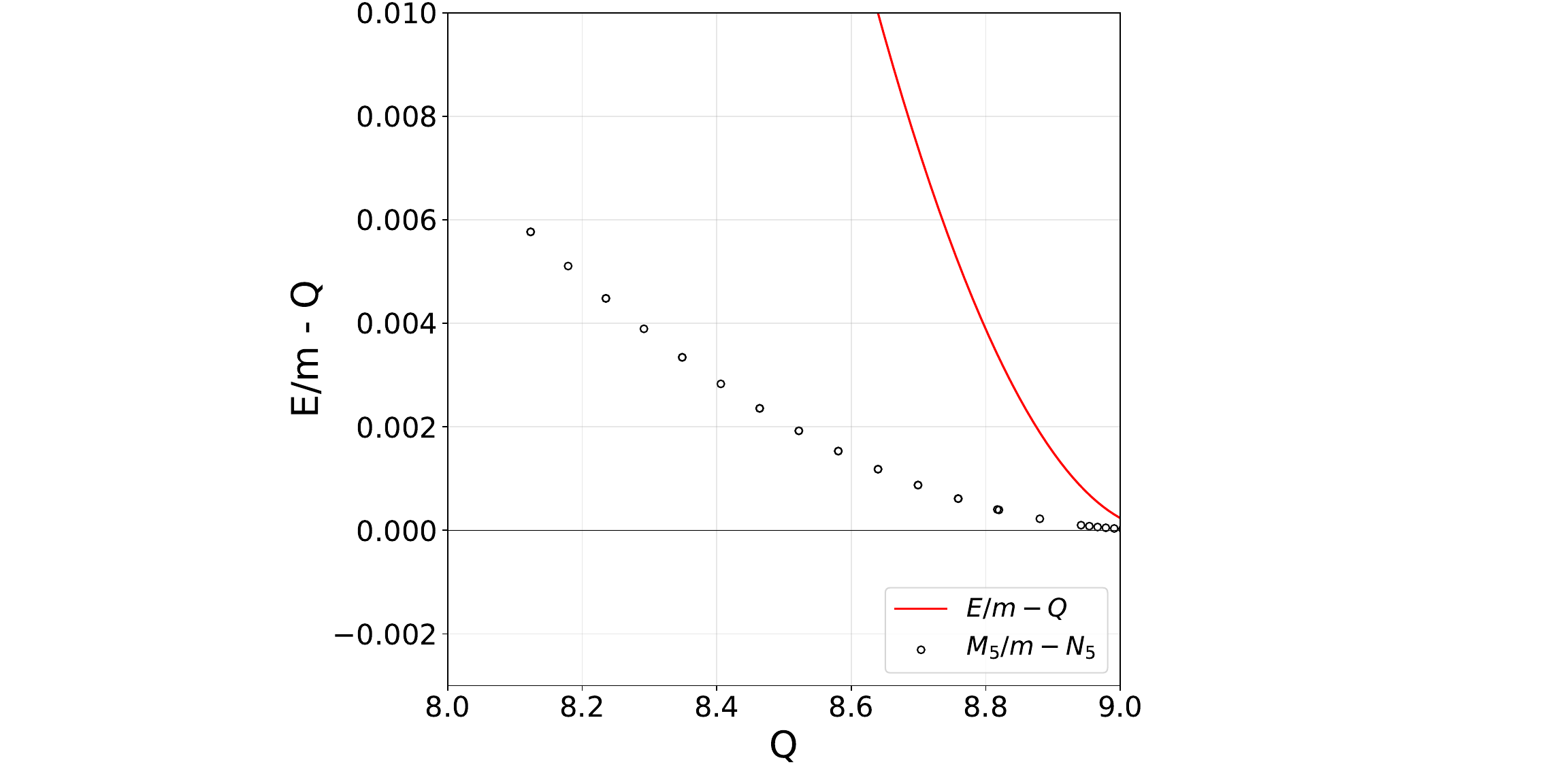}
    \captionof{figure}{Plot of the configuration energy versus the $U(1)$ charge for GR (circles) and the Wick-Cutkosky model (red line).}
    \label{Fig:5}
\end{center}
The small deviation (i.e., incapable of stabilizing the system and shifting the values of $(E/m)-Q$ into the negative region) of the $M_5(N_5)$ dependence from $E(Q)$ in Figs. \ref{Fig:4} and \ref{Fig:5} is determined by the pattern of weak scale-invariance breaking caused by the introduction of next-order corrections. The difference is more noticeable for the $E(\omega)$ and $Q(\omega)$ dependencies, which indicates the importance of taking into account corrections not only for the complex scalar field but also for gravity. Moreover, Fig. \ref{Fig:5} implies the kinematic instability of the obtained boson star type solution. The violation of scale invariance led to the lifting of degeneracy and to an instability associated with the appearance of a decay channel.

We also construct profiles for the complex scalar field for the full GR theory (i.e., the solution \eqref{eq:3.13}) and for the five-dimensional generalization of the Wick--Cutkosky model at different values of $\kappa_c$ ($\bar{v}^2$). Similarly, profiles of the metric functions will be constructed.
\begin{center}
    \includegraphics[width=\textwidth, trim=0 2cm 0 3cm, clip]{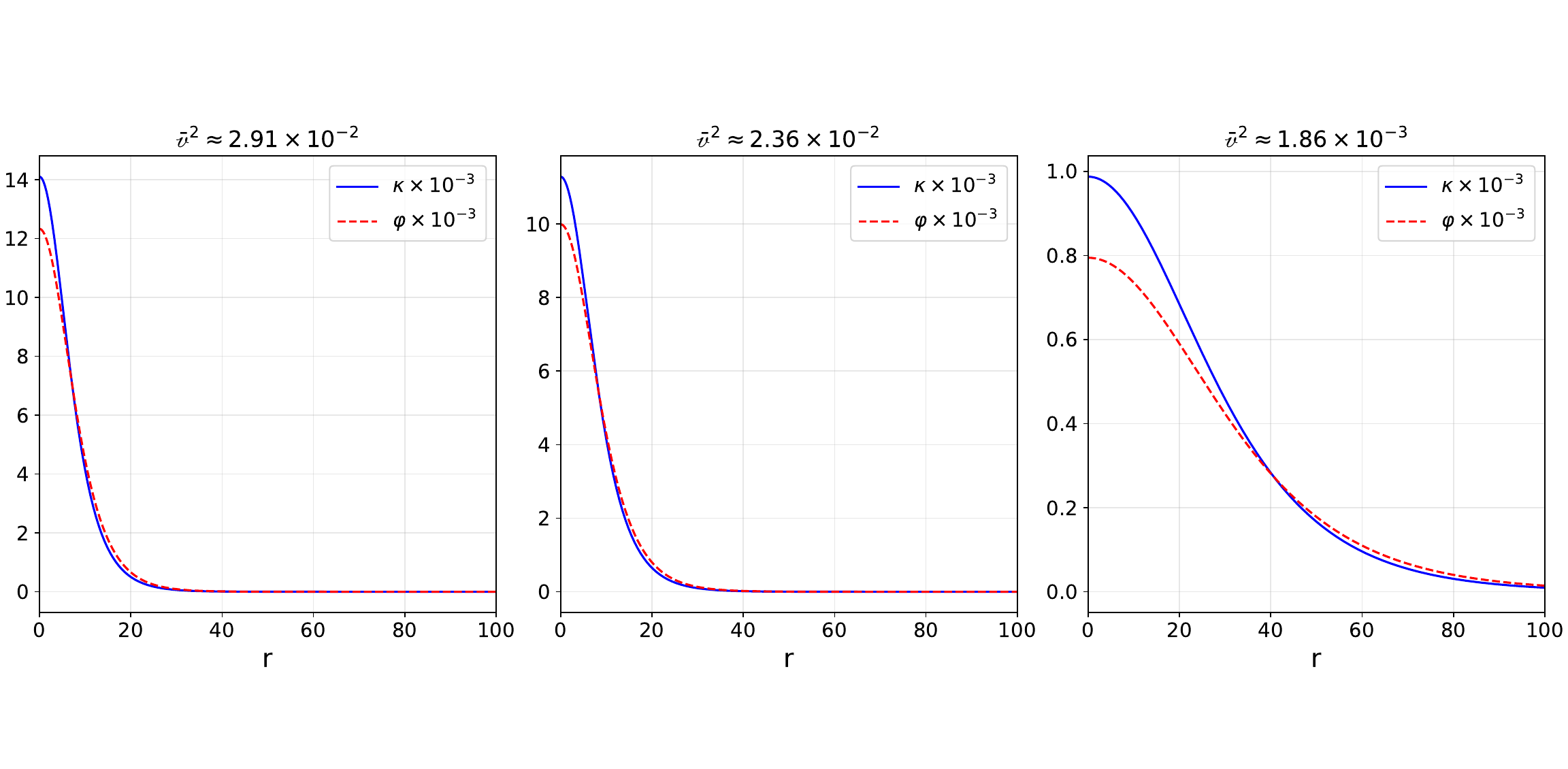}
    \captionof{figure}{Plots of the scalar field profiles in GR and the Wick--Cutkosky model for various values of $\bar{v}^2$. The red dashed line corresponds to the Wick--Cutkosky model, and the blue solid line is the profile for GR.}
    \label{Fig:6}
\end{center}
\begin{center}
    \includegraphics[width=\textwidth, trim=0 2cm 0 3cm, clip]{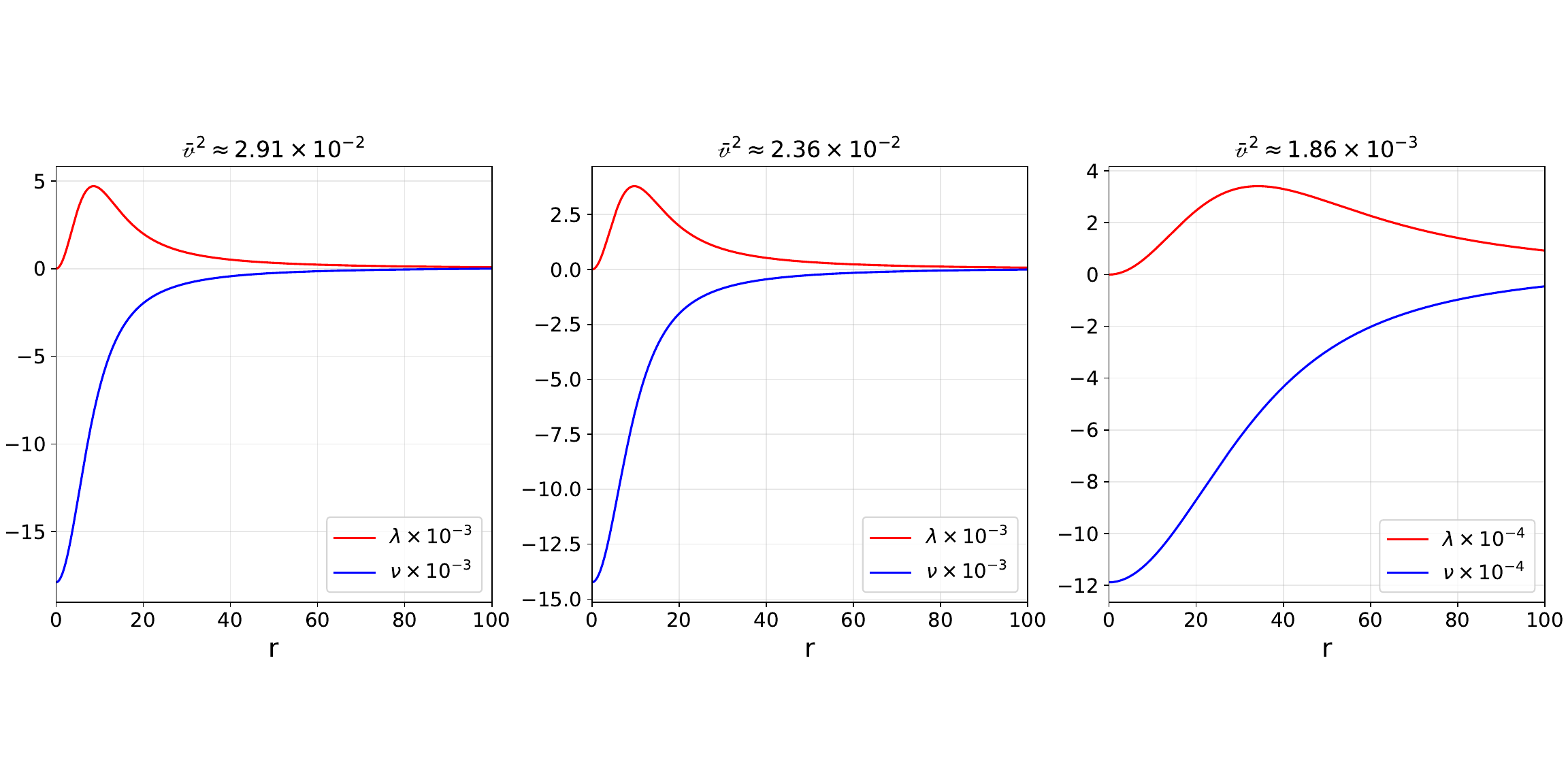}
    \captionof{figure}{Plots of the metric function profiles for various values of $\bar{v}^2$. The red solid line corresponds to the profile $\lambda(r)$, and the blue one to $\nu(r)$.}
    \label{Fig:7}
\end{center}
From the appearance of Figs. \ref{Fig:6} and \ref{Fig:7}, one can conclude that including corrections of the next order of smallness makes a significant contribution, not only for the complex scalar field but also for gravity. They must be taken into account. As $\bar{v}^2$ approaches zero, the amplitudes decrease.

\section{Further remarks and conclusions}
In four spatial dimensions, the Newtonian Schrödinger–Poisson system is scale invariant. Consequently, its localized boson star solutions form a one-parameter family related by dilatations, while the dimensionless energy/binding-energy functional is degenerate. In this work we confirmed that relativistic corrections explicitly break this symmetry and lift the degeneracy, producing the behavior observed in the full five-dimensional Einstein–Klein–Gordon solutions.

The feature in the pattern of dynamical evolutions of minimal (4+1)-dimensional boson stars  is related to the vanishing of the binding energy at a finite maximum particle number \cite{marks2026}, it is associated to  
the restoration of scale invariance of the system in the Newtonian (non-relativistic) limit. As was shown in Sec. 2, the degeneracy of the $M_5-mN_5$ spectrum into the point $\text{H}=0$ is a direct consequence of the conservation of the dilatation current.

The considered theory (see Sec. 1 and Sec. 3) is naively similar in structure to the (4+1)-dimensional generalization of the Wick--Cutkosky model. This circumstance, together with the weak violation of scale symmetry by relativistic corrections for the complex scalar field, allows a qualitative understanding of the cause of the instability of minimal BSs. The inclusion of new terms in the Lagrangian leads to a loss of symmetry and removal of degeneracy. The system passes from the critical point to an unstable branch of solutions. Accounting for post-Newtonian (PN) corrections will lead to a shift of the configuration from the unstable branch corresponding to the Wick--Cutkosky model down to the energy of free particles and bound states. However, due to the specific features of (4+1) dimensional spacetime, gravity alone will not be able to balance the repulsion and stabilize the system, as reflected in Fig.~\ref{Fig:4}. This circumstance complicates the formation of massive relativistic BSs in various mechanisms \cite{Kasuya:2025puk,Li:2026yhv}, although the introduction of self-interaction potentials may improve the situation.

Finally, it can also be noted that when introducing corrections of the next order in the small parameter, it is important to take into account the contribution of gravity, and not only that of antiparticles. In the case considered in Section 1, it was shown that taking the limit $\frac{m}{M_{5Pl}}\rightarrow 0$ causes difficulties, since it leads to the elimination of the interaction of bosons with gravity. This argument in favor of the comparability of PN- and relativistic corrections is confirmed by the numerical results shown in Figures \ref{Fig:4}--\ref{Fig:7}. Also, NR corrections could be crucial for the consideration of linear perturbations on the compact field lumps \cite{Evslin:2026eph}.
Taking into account next to the leading orders of PN-corrections for extended objects is highly demanding and was discussed in detailed review \cite{Goldberger:2004jt}.

In this
paper, we have discussed only spherically symmetric  configurations parametrized by the ansatz \eqref{eq:18}. 
An analysis of the non-relativistic limit of the more general case of rotating boson stars in five dimensions \cite{Hartmann:2010pm} appears to be of interest for further research.

The work here may be taken further by considering the non-relativistic limit of 4+1-dimensional black holes with scalar hair. We hope that we can
address these issues in our future work. 

\section*{Acknowledgments}
The authors express their gratitude to Mikhail Smolyakov and Ivan Ivashkin for their helpful comments and discussions.
The work of E. Kim was supported by the Foundation for the Advancement of Theoretical Physics and Mathematics BASIS. 

This work is dedicated to the memory of Dmitry Levkov, an outstanding physicist and a very close friend of us. The results presented here originate from our numerous discussions with him.

\appendix
\section{Integral characteristics}\label{Integral characteristics}
In this appendix, we define some integral characteristics that are used throughout this work. In particular, the conserved $U(1)$ charge, associated with the invariance of \eqref{eq:19} under transformation $\psi \rightarrow e^{i\alpha}\psi$. The corresponding conserved current can be found using Noether's method by introducing local dependence of small parameter $\alpha$ on spacetime coordinates $x^\mu$. On equations of motion
\[
\delta S_{NR}=0=\int d^5x\Bigl(i\partial_0(i\alpha\psi)\psi^*+i\partial_0\psi(-i\alpha\psi^*)-\frac{1}{2m}\nabla(i\alpha\psi)\nabla\psi^*-\frac{1}{2m}\nabla\psi\nabla(-i\alpha\psi^*)\Bigl)=
\]
\[
=\int d^5x\Bigl(-\partial_0\alpha\psi\psi^*-\frac{i}{2m}\nabla\alpha\psi\nabla\psi^*+\frac{i}{2m}\nabla\alpha\psi^*\nabla\psi\Bigl)=
\]
\[
=-\int d^5x\Bigl[\partial_0\alpha\Bigl(\psi\psi^*\Bigl)\Bigr]-\int d^5x\Bigl[\nabla\alpha\frac{i}{2m}\Bigl(\psi\nabla\psi^*-\psi^*\nabla\psi\Bigl)\Bigr].
\]
The last equality immediately leads to the components of the conserved current:
\[
j^0=\psi\psi^*,\qquad j^i=-\frac{i}{2m}(\psi^*\nabla\psi-\psi\nabla\psi^*).
\]
The conservation of the Noether current, expressed by $\partial_\mu j^\mu=0$, determines the conserved quantity of the system - the $U(1)$ charge:
\begin{equation}
    N=\int d^4xj^0=\int d^4x\psi\psi^*=2\pi^2\int r^3\phi^2(r)dr.
    \label{eq:2.1}
\end{equation}
When studying the weak violating of scale symmetry, the contribution from including gravitational and relativistic corrections is investigated using the dependence of the binding energy on the $U(1)$ charge of the system. To this end, it is necessary to determine the mass of the configuration $M_5$ and the $U(1)$ charge $N_5$ in the full GR theory
\begin{equation}
    M_5=\lim_{r\rightarrow\infty}\left[\frac{3\pi r^2}{8\Gamma(2)}\left(1-e^{-\lambda}\right)\right];
\end{equation}
\begin{equation}
    N_5=\frac{\pi}{4\Gamma(2)}\omega\int dr\Bigl(r^3e^{\frac{\lambda-\nu}{2}}\kappa^2\Bigr).
\end{equation}
In determining these, we followed the notation adopted in \cite{franzin2024}.

\section{General Relativity Equations in $(4+1)$}\label{appendix a}
Here the basic relations of general relativity in five-dimensional spacetime are presented: the Ricci tensor and the metric energy-momentum tensor:
\[
    \Gamma^\mu_{\nu\lambda}=\frac{1}{2}g^{\mu\rho}[\partial_\lambda g_{\rho\nu}+\partial_\nu g_{\rho\lambda}-\partial_\rho g_{\nu\lambda}];
\]
\begin{equation}
R_{\sigma\nu}=R^\rho_{\sigma\rho\nu}=\partial_\rho\Gamma^\rho_{\nu\sigma}-\partial_\nu\Gamma^\rho_{\rho\sigma}+\Gamma^\rho_{\rho\lambda}\Gamma^\lambda_{\nu\sigma}-\Gamma^\rho_{\nu\lambda}\Gamma^\lambda_{\rho\sigma};
\label{eq:3}
\end{equation}
\begin{equation}
    T_{\mu\nu}=-\frac{2}{\sqrt{-g}}\frac{\delta \mathcal{L}_M}{\delta g^{\mu\nu}},
    \label{eq:4}
\end{equation}
where this corresponds to the first two terms in \eqref{eq:1}. We note that expression for $R_{\mu\nu}$ differs from the analogous one in \cite{Weinberg:1972}.

The Einstein equations corresponding to \eqref{eq:1} are also given:
\begin{equation}
R_{\mu\nu}=8\pi G_5\Bigl(T_{\mu\nu}-\frac{1}{3}g_{\mu\nu}T_\rho^\rho\Bigr).
\label{eq:2}
\end{equation}
This expression differs by the coefficient $\frac{1}{3}$ from $\frac{1}{2}$ in front of the second term on the right-hand side in the $(3+1)$-dimensional theory.

Finally, we indicate the orders of derivatives that follow directly from the characteristic scales of the system
\[
\partial_i\sim\frac{1}{\bar{r}},\qquad\partial_0\sim\frac{\bar{v}}{\bar{r}}.
\]

\bibliography{references}

\end{document}